\documentclass[aps,prl,floatfix,twocolumn,reprint,amsmath,amssymb,superscriptaddress]{revtex4-1}
\usepackage{amsfonts}
\usepackage{mathrsfs}
\usepackage{amsmath}
\usepackage{color}
\usepackage{natbib}
\usepackage{graphicx}
\usepackage{bm}
\usepackage{amssymb}
\usepackage{xspace}
\usepackage{epstopdf}
\usepackage{dcolumn}
\usepackage{longtable}
\usepackage{multirow}
\usepackage[colorlinks=true, letterpaper=true, pdfstartview=FitV, linkcolor=blue, citecolor=blue, urlcolor=blue]{hyperref}

\makeatletter

\newcommand{\Rmnum}[1]{\expandafter\@slowromancap\romannumeral #1@}
\makeatother

\begin{document}
\title{Type-III Weyl Semimetals: $\mathrm{(\mathrm{Ta}Se_{4})_{2}I}$}

\author{Xiao-Ping Li}
\thanks{These authors contributed equally to this work.}
\affiliation{Key Lab of advanced optoelectronic quantum architecture and measurement (MOE), Beijing Key Lab of Nanophotonics $\&$ Ultrafine Optoelectronic Systems, and School of Physics, Beijing Institute of Technology, Beijing 100081, China}

\author{Ke Deng}
\thanks{These authors contributed equally to this work.}
\affiliation{State Key Laboratory of Low Dimensional Quantum Physics and Department of Physics, Tsinghua University, Beijing 100084, China}

\author{Botao Fu}
\thanks{These authors contributed equally to this work.}
\affiliation{College of Physics and Electronic Engineering, Center for Computational Sciences, Sichuan Normal University, Chengdu, 610068, China}
\affiliation{Key Lab of advanced optoelectronic quantum architecture and measurement (MOE), Beijing Key Lab of Nanophotonics $\&$ Ultrafine Optoelectronic Systems, and School of Physics, Beijing Institute of Technology, Beijing 100081, China}

\author{YongKai Li}
\thanks{These authors contributed equally to this work.}
\affiliation{Key Lab of advanced optoelectronic quantum architecture and measurement (MOE), Beijing Key Lab of Nanophotonics $\&$ Ultrafine Optoelectronic Systems, and School of Physics, Beijing Institute of Technology, Beijing 100081, China}

\author{Da-Shuai Ma}
\affiliation{Key Lab of advanced optoelectronic quantum architecture and measurement (MOE), Beijing Key Lab of Nanophotonics $\&$ Ultrafine Optoelectronic Systems, and School of Physics, Beijing Institute of Technology, Beijing 100081, China}

\author{JunFeng Han}
\affiliation{Key Lab of advanced optoelectronic quantum architecture and measurement (MOE), Beijing Key Lab of Nanophotonics $\&$ Ultrafine Optoelectronic Systems, and School of Physics, Beijing Institute of Technology, Beijing 100081, China}

\author{Jianhui Zhou}
\email{jhzhou@hmfl.ac.cn}
\affiliation{Anhui Province Key Laboratory of Condensed Matter Physics at Extreme Conditions, High Magnetic Field Laboratory, Chinese Academy of Sciences (CAS), Hefei 230031, Anhui, China}

\author{Shuyun Zhou}
\affiliation{State Key Laboratory of Low Dimensional Quantum Physics and Department of Physics, Tsinghua University, Beijing 100084, China}
\affiliation{Frontier Science Center for Quantum Information, Beijing, China}

\author{Yugui Yao}
\email{ygyao@bit.edu.cn}
\affiliation{Key Lab of advanced optoelectronic quantum architecture and measurement (MOE), Beijing Key Lab of Nanophotonics $\&$ Ultrafine Optoelectronic Systems, and School of Physics, Beijing Institute of Technology, Beijing 100081, China}

\begin{abstract}
Weyl semimetals have been classified into type-I and type-II with respect to the geometry of their Fermi surfaces at the Weyl points.
Here, we propose a new class of Weyl semimetal, whose unique Fermi surface contains two electron or two hole pockets touching at a multi-Weyl point, dubbed as type-III Weyl semimetal.
Based on first-principles calculations, we first show that quasi-one-dimensional compound $\mathrm{(\mathrm{Ta}Se_{4})_{2}I}$ is a type-III Weyl semimetal with larger chiral charges.
$\mathrm{(\mathrm{Ta}Se_{4})_{2}I}$ can support four-fold helicoidal surface states with 
 long Fermi arcs on the (001) surface.
 Angle-resolved photoemission spectroscopy measurements are in agreement with the gapless nature of (TaSe4)2I at room temperature and reveal its characteristic dispersion. In addition, our calculations show that external strain could induce 
transitions in $\mathrm{(\mathrm{Ta}Se_{4})_{2}I}$ among the type-III, type-II, and type-I Weyl semimetals, accompanied with the Lifshitz transitions of the Fermi surfaces.
Therefore, our work first experimentally indicates $\mathrm{(\mathrm{Ta}Se_{4})_{2}I}$ as a type-III Weyl semimetal and provides a promising platform to further investigate the novel physics of type-III Weyl fermions.
\end{abstract}
\maketitle
%
\textit{Introduction.--}Recently, three-dimensional (3D) Weyl/Dirac semimetals possessing discrete and finite degenerate points, Weyl points (WPs), in the Brillouin zone (BZ), have attracted increasing attentions~\cite{Hosur13Physique,Bansil2016RMP,Burkov2016Nmat,armitage2018rmp}.
Weyl/Dirac semimetals have been predicted in a large number of materials~\cite{Zhang2019Nature,Tang2019Nature,Vergniory2019Nature} and also exhibit various novel physical phenomena such as the chiral magnetic effect, ultrahigh mobility, negative longitudinal magnetoresistance and 3D quantum Hall effect, some of which have been recently confirmed experimentally~\cite{YanBH2017arcmp,ZHANG2018SciB,armitage2018rmp}.
It is known that Weyl semimetals can be characterized by the chiral charge of the WP: single Weyl semimetals with chiral charge $\chi=\pm1$~\cite{murakami2007njp,wan2011prb,Young2012PRL} and multi-Weyl (double or triple) semimetals with larger chiral charges $\chi=\pm2,\pm3$ and the multi-fold Fermi arc states~\cite{xu2011prl,fang2012prl,tsirkin2017composite,huang2016pnas}.
The unique nonlinear energy dispersions of multi-Weyl semimetals may lead to striking non-Fermi-liquid behaviors~\cite{HanSE2019PRL,WangJR2019PRB,zhang2018arxiv}.
However, to date, only the single Weyl/Dirac semimetals have been observed in experiments, and it remains important to search for realistic materials for multi-Weyl semimetals.

\begin{figure}[h]
\includegraphics[width=8.2cm]{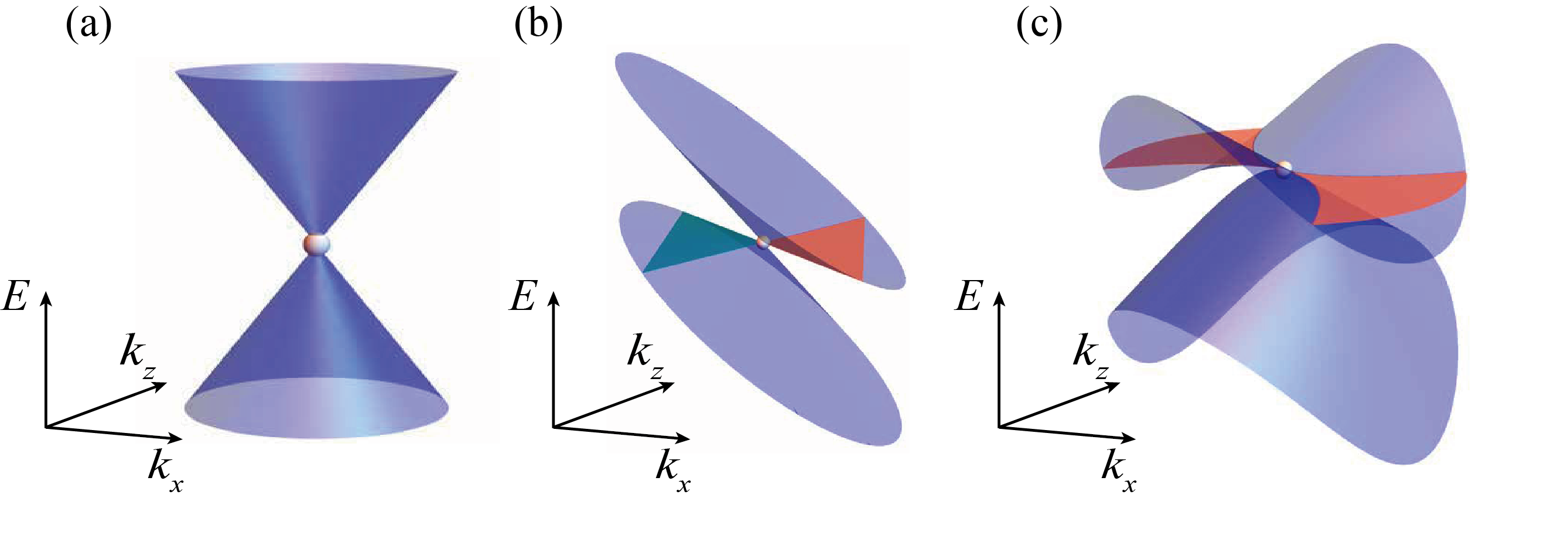}
\caption{(Color online) Schematic for three types of Weyl semimetals.
(a) and (b) show the type-I and type-II Weyl semimetals with point-like and contacted electron and hole pockets, respectively.
(c) Fermi surface of type-III Weyl semimetal has two contacted electron pockets (two hole pockets are for the opposite quadratic tilting term).
\label{fig1}}
\end{figure}

The Fermi surface plays a crucial rule in understanding the fundamental physics of crystals, such as superconductivity, the charge/spin density wave~\cite{Anderson1984BCCMP} and anomalous transport~\cite{xiao2010rmp,Nagaosa2010RMP}.
According to the geometry of the Fermi surfaces at the WPs, Weyl semimetals can also been classified into type-I and type-II~\cite{Soluyanov2015nature}.
Unlike the point-like Fermi surface of type-I Weyl semimetals~(Fig.~\ref{fig1}(a)),
the type-II Weyl semimetals with over-tilted Weyl cones~(Fig.~\ref{fig1}(b))~\cite{XuPRL2015,deng2016NP,huang2016NM,Tamai2016PRX,jiang2017NC,Yao2019PRL}, whose Fermi surface consists of touched electron-hole pockets, break the Lorentz invariance and lead to many unusual electromagnetic responses~\cite{Soluyanov2015nature,Yuzm2016PRL,Tchoumakov2016PRL,Udagawa2016PRL,Obrien2016prl,chang2017prl}.
For example, magneto-transport experiment in $\mathrm{\mathrm{W}Te_{2}}$ shows that the tilting of Weyl cone makes the chiral anomaly (origin of various intriguing physical effects in Weyl semimetals), and the resultant negative longitudinal magnetoresistance exhibits strong orientation dependence and even disappearance~\cite{Wang2016NC}.
Thus, it is fundamentally important whether there exists new Weyl semimetal whose Fermi surface is distinctly different from the counterparts of type-I and -II Weyl semimetals, such as,
 two contacted electron or two hole pockets at the WP (Fig.~\ref{fig1}(c)). If so, what are the relevant chiral charge and the unique physical properties?

In this work, we first discover a new type-III Weyl semimetal whose Fermi surface consists of two electron or two hole pockets touching at the WPs
in the quasi-one dimensional (1D) compound $\mathrm{(\mathrm{Ta}Se_{4})_{2}I}$.
$\mathrm{(\mathrm{Ta}Se_{4})_{2}I}$ possesses two pairs of double WPs, thus supporting the four-fold helicoidal surface states on (001) surface with 
long Fermi arcs.
Meanwhile, strains that break $C_4$ symmetry could trigger transitions from type-III WPs to type-II and type-I ones, accompanied with the Lifshitz transitions of the Fermi surfaces. More importantly, our angle-resolved photoemission spectroscopy (ARPES) results suggest $\mathrm{(\mathrm{Ta}Se_{4})_{2}I}$ is a type-III Weyl semimetal and provides the first material realization of multi-Weyl semimetals.

\textit{Type-III Weyl fermions.--}To illustrate the essential feature of the type-III Weyl fermions, we consider a general Hamiltonian near the WP,
\begin{equation}\label{genericH}
\mathcal{H}_{\text{n}}(\bm{k})=w_{z}k_{z}+w_{\parallel}k_{\parallel}^{2}+v_{z}k_{z}\sigma_{z}+(ak_{\pm}^{n}\sigma_{+}+h.c.),
\end{equation}
and the corresponding eigenvalues are $E_{\pm}(\bm{k})=w_{z}k_{z}+w_{\parallel}k_{\parallel}^{2} \pm \sqrt{v_{z} ^2 k_{z}^2+a^2 k_{\parallel}^{2 n}}$.
Here $k_{\parallel}=\sqrt{k_x^2+k_y^2}$, $k_{\pm}=k_x\pm ik_y$, $\sigma_\pm=\frac{1}{2}(\sigma_x\pm i\sigma_y)$, $\sigma_{x,y,z}$ are the Pauli matrices.
The subscript $n=1,2,3$ denotes the single-, double- and triple-WPs, respectively.
The first two terms refer to the linear and quadratic tilting terms, which are crucial for the emergence of type-II and type-III Weyl fermions, respectively.
To be specific, the linear tilting term $w_z k_z$ (e.g. $w_z>0$) pulls down the band for $k_z<0$ and pushes up the band for $k_z>0$.
Once $|w_z|>|v_z|$, the Weyl cone~(Fig.~\ref{fig1}(a)) is over tilted and the Fermi surface changes from a point to a coexistence of electron and hole pockets, leading to a type-II WP~(Fig.~\ref{fig1}(b))~\cite{Soluyanov2015nature}.
On the other hand, the quadratic tilting term $w_{\parallel} k_{\parallel}^{2}$ ($w_{\parallel}>0$) always pushes up energy bands for any in-plane $k$-path ($k_{z}$=0).
It is worth noting that the quadratic tilting term plays an important role in multi-Weyl semimetals. Let us first focus on the double Weyl semimetals.
When $w_{\parallel} k_{\parallel}^{2}$ dominates the energy dispersion in the $k_x$-$k_y$ plane ($|w_{\parallel}|>|a|$), the Weyl cone can be over-tilted, giving rise to a type-III WP.
The resulting Fermi surface consists of two contacted electron or two hole pockets (Fig.~\ref{fig1}(c)).
The sign of the quadratic tilting term $w_{\parallel}$ determines the types of the two contacted pockets.
Our classification of Weyl fermions can be well understood in terms of the general titling mechanism. Specifically, the dominant $n$th order tilting term would lead to the type-$(n+1)$ Weyl fermions~\cite{SMtiii}. It should be emphasized that the type-III WP can emerge in triple Weyl semimetals as well.
%

\begin{figure}[h]
\includegraphics[width=8.2cm]{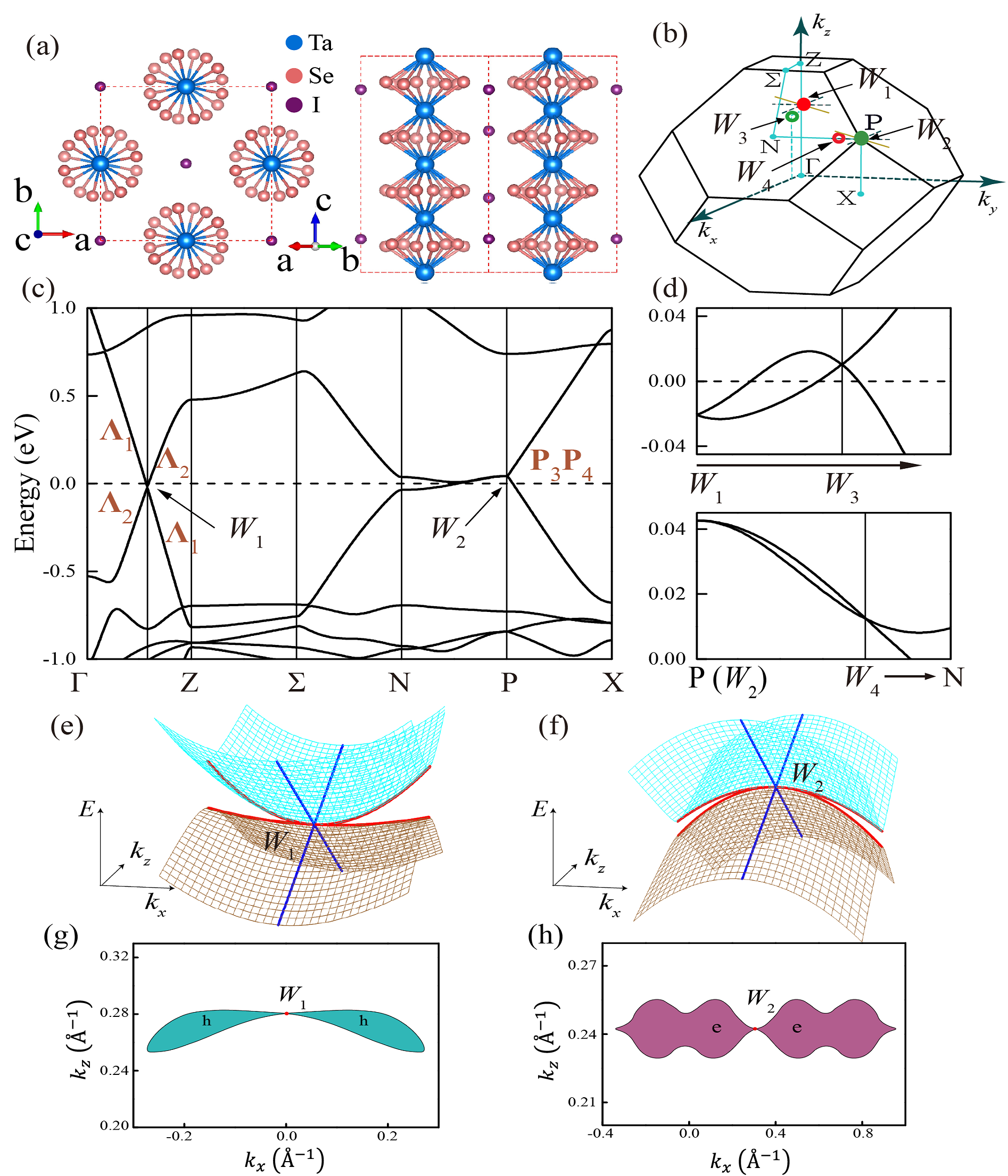}
\caption{(Color online)
(a) Top view and side view of crystal structure of $\mathrm{(\mathrm{Ta}Se_{4})_{2}I}$.
(b) The high symmetry points $\Gamma$, X, P, Z, N and the WPs $W_{1,2,3,4}$ are marked in the first BZ.
(c) The electronic structure of $\mathrm{(\mathrm{Ta}Se_{4})_{2}I}$ along a selected path.
$\Lambda_{1,2}$ and $\mathrm{P}_{3,4}$ label the irreducible representations of two crossing bands.
(d) The zoomed band structures along the P-N path and $W_1$-$W_3$ path.
(e)-(f) 3D plots of the energy dispersions near WPs $W_{1,2}$ in the $k_{x}$-$k_{z}$ plane. The red and blue lines indicate the dispersions along the $k_{x}$ and $k_{z}$ directions, respectively.
(g)-(h) The calculated constant energy contours in $k_{x}$-$k_{z}$ plane at the energy of $W_{1,2}$ points, respectively. The cyan zone stands for the electron pocket and the purple zone for the hole pocket.~\label{fig2}}
\end{figure}

\textit{Material realization.--}$\mathrm{(\mathrm{Ta}Se_{4})_{2}I}$ is a typical quasi-1D compound that has been synthesized more than thirty years ago~\cite{gressier1982preparation}. It forms a body-centred tetragonal lattice with the chiral space group $I422$ at room temperature~\cite{gressier1984characterization,chang2018NM}. As shown in Figs.~\ref{fig2}(a)-\ref{fig2}(b), crystal of $\mathrm{(\mathrm{Ta}Se_{4})_{2}I}$ contains right-handed TaSe$_4$ chains along the $c$-axis with halogen atoms filling inter-chain interstitial regions.
Very recently, the charge density wave (CDW) phase~\cite{wang1983zz,maki1983charge,tournier2013PRL} and possible axion physics in $\mathrm{(\mathrm{Ta}Se_{4})_{2}I}$ have generated a lot of interest~\cite{WangZ13prb,gooth2019arxiv,shi2019axioncdw}.

The electronic structure of $\mathrm{(\mathrm{Ta}Se_{4})_{2}I}$ in Fig.~\ref{fig2}(c) displays several notable features~\cite{SOC}.
First, we find that a band crossing ($W_{1}$) appears along the $C_4$ invariant ${\Gamma}$-Z path with linear dispersion in a broad energy range from -0.5 eV to 0.5 eV.
Another band crossing emerges at the corner of the BZ ($W_{2}$), which also has linear dispersion along the P-X path.
Second, two bands along the P-N path exhibit weak dispersions close to the Fermi level, which is attributed to the weak inter-chain coupling in $\mathrm{(\mathrm{Ta}Se_{4})_{2}I}$.
Third, the zoom region in Fig.~\ref{fig2}(d) shows that these two bands stemming from P point will switch their orders and create a WP ($W_{4}$) with over-tilted linear dispersion.
Similar band switching process happens near $W_{1}$ point, forming another WP at $W_{3}$ in the $k_y=0$ plane.
That is, there exist four independent WPs marked as $W_{1,2,3,4}$ in the irreducible BZ.
Combining crystal symmetries with time reversal symmetry, we obtain two $W_{1,2}$ points and eight $W_{3,4}$ points in total in the first BZ.

Figs.~\ref{fig2}(e)-\ref{fig2}(f) show that the WPs $W_{1}$ and $W_{2}$ have unusual quadratic dispersion along the $k_{x}$ path and linear dispersion along the $k_{z}$ path, indicating larger chiral charges.
To verify this point, we calculate their chiral charge and find $\chi$=-2(2) for $W_{1}$($W_{2}$), $\chi$=1(-1) for $W_{3}$($W_{4}$), respectively.
Moreover, the quadratic tilting terms make the valence band over-titled along the $k_{x}$ direction for $W_{1}$ and the conduction band over-titled around the $k_{x}$ direction for $W_{2}$, leading to unique Fermi surfaces. As shown in Figs.~\ref{fig2}(g)-\ref{fig2}(h), the resulting Fermi surface near $W_{1}$ ($W_{2}$) contains two contacted hole (electron) pockets.
This is the remarkable signature of the aforementioned type-III Weyl fermions.


\begin{table}[t]
\caption{Parameters of effective Hamiltonians in Eq.~(\ref{Ht3}) and Eq.~(\ref{Hdw}) for the double WPs $W_{1}$ and $W_{2}$, respectively. $a_i$, $b_i$, $c_i$, $d_i$ are fitting parameters that are determined by the first-principles calculations~\cite{SMtiii}.}
\vspace{0.5cm}
\begin{tabular}{|c|c|l|l|}
\hline

$WP$& $\chi$ & \multicolumn{1}{c|}{\begin{tabular}[c]{@{}c@{}} $H_{DW}$\end{tabular}} & \multicolumn{1}{c|}{\begin{tabular}[c]{@{}c@{}} $\omega_{t}$\end{tabular}} \\ \hline

$W_{1}$ & $-2$ & \begin{tabular}[c]{@{}l@{}}$\omega_{x}^{(1)}=a(k_{x}^{2}-k_{y}^{2})+a_{1}k_{x}k_{y}$\\ $\omega_{y}^{(1)}=b(k_{x}^{2}-k_{y}^{2})+b_{1}k_{x}k_{y}$\\ $\omega_{z}^{(1)}=c_{1}k_{z}+c_{2}(k_{x}^{2}+k_{y}^{2})$\end{tabular} & 
\begin{tabular}[c]{@{}l@{}}$\omega_{t}^{(1)}=d_{1}k_{z}+d_{2}(k_{x}^{2}+k_{y}^{2})$\end{tabular} \\ \hline

$W_{2}$  & $2$  & \begin{tabular}[c]{@{}l@{}}$\omega_{x}^{(2)}=a_{1}k_{z}+a_{2}k_{x}k_{y}$
 \\ $\omega_{y}^{(2)}=b_{1}k_{z}+b_{2}k_{x}k_{y}$\\ $\omega_{z}^{(2)}=c_{1}(k_{x}^{2}-k_{y}^{2})$\end{tabular} & $\omega_{t}^{(2)}=d_{1}(k_{x}^{2}+k_{y}^{2})$  \\ \hline
\end{tabular}
\end{table}

\textit{Effective ${k\cdot p}$ models.--}To gain more insights into Weyl fermions in $\mathrm{(\mathrm{Ta}Se_{4})_{2}I}$, we would like to build the effective ${k\cdot p}$ model based on symmetry analysis.
Along the ${\Gamma}$-Z path, the corresponding little group is $C_4$ group.
Two bands that cross the Fermi level host two distinctive 1D irreducible representations, $\Lambda_1$ and $\Lambda_2$~\cite{elcoro2017double}, the corresponding eigenvalues of $C_4$ are $\pm1$.
It implies that a double WP ($W_1$) can appear once band inversion happens~\cite{fang2012prl}, as shown in Fig.~\ref{fig2}(c).

The P point at the corner of BZ has the little group of $D_2$, which possesses four non-equivalent 1D representations that do not guarantee two-fold degeneracy.
In reality, the P point is invariant under a joint operation of four-fold rotation and time reversal operation ($C_{4}T$).
According to Herring rules~\cite{Herring1937PR}, two bands around the Fermi level with representations P$_3$ and P$_4$ can stick together, forming a WP at the high symmetry point.
By further taking into account the constraints from other crystal symmetries, the energy dispersion of WP at P ($W_2$) point is quadratic in the $k_{x}$-$k_{y}$ plane, forming a double WP.
Finally, we obtain the effective ${k\cdot p}$ Hamiltonian of double WPs, $W_{1}$ and $W_{2}$ as follows (see Supplemental Material~\cite{SMtiii}),
\begin{eqnarray}\label{Ht3}
H_{III}^{(i)}=H_{DW}^{(i)}+w_{t}^{(i)},\label{HGt}
\end{eqnarray}
where $H_{DW} ^{(i)}$ describes a type-I double Weyl fermion that has the form
\begin{eqnarray}\label{Hdw}
H_{DW}^{(i)}&	=&w_{x}^{(i)}\sigma_{x}+w_{y}^{(i)}\sigma_{y}+w_{z}^{(i)}\sigma_{z},\label{HGI}
\end{eqnarray}
and $w_{t}^{(i)}$ is the tilting term. $i=1,2$ refers to $W_i$.
The explicit expressions $w_{x,y,z}^{(i)}$ and $w_{t}^{(i)}$ are given in Table-I.

%
One can see that the $w_{t}^{(1)}$ includes two parts: the linear and quadratic tilting terms.
Based on our theoretical calculations, the linear term along the $k_z$ direction is tiny but the quadratic term along the $k_x$ or $k_y$ direction is dominating.
The $w_{t}^{(2)}$ only contains the quadratic term due to the symmetry constraints.
It has been pointed out that the over-tilted linear term can induce type-II Weyl semimetals with touching electron-hole pockets, while the over-tilted quadratic term plays a key role in forming type-III Weyl semimetals.
Specifically, $w_{t}^{(1)}$ pushes the valence band around $W_1$ upward along the $k_{x}$ or $k_{y}$ path.
When $|d_{2}|>\sqrt{a^{2}+b^{2}+c_{2}^{2}}$, we obtain a type-III WP with two contacted hole pockets (Fig.~\ref{fig2}(g)).
Similarly, $w_{t}^{(2)}$ pulls the conduction band around $W_2$ downward along the $k_{x}$ or $k_{y}$ path when $|d{}_{1}|>|c_{1}|$ and gives birth to another type-III Weyl fermion with two contacted electron pockets (Fig.~\ref{fig2}(f)).
These coefficients of ${k\cdot p}$ Hamiltonian obtained from fitting with our calculations confirm these analysis (see Supplemental Material~\cite{SMtiii}).

\begin{figure}[t]
\includegraphics[width=8.2cm]{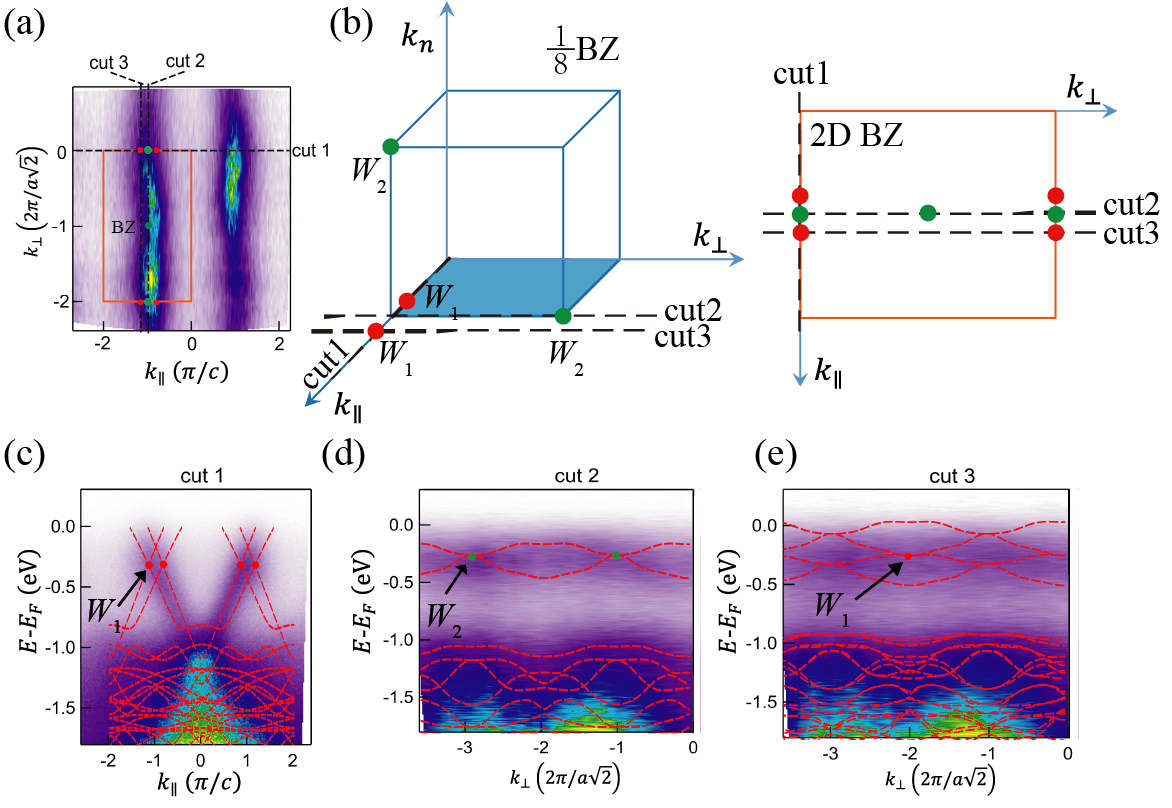}
\caption{(Color online) ARPES spectra of (TaSe$_4$)$_2$I on (110) surface and schematics.
The $k_\parallel$ ($k_\perp$) denotes the wave vector along the direction parallel (perpendicular) to the direction of TaSe$_4$ chain.
(a) The constant energy contour map is measured at Fermi level. The red dashed rectangle indicates the surface BZ.
(b) An schematic illustration of the bulk and surface BZ of conventional cell of (TaSe$_4$)$_2$I. Three cuts (1, 2, 3) marked by black dashed lines pass through $W_1$ or $W_2$.
(c)-(e) ARPES spectra measured along cuts (1, 2, 3) at photon energy of 21.2 eV, in comparison with the calculated band structures (red dashed lines).
\label{fig3}}
\end{figure}
%

\textit{ARPES spectra and helicoidal surface states.--}Fig.~\ref{fig3} shows the ARPES measurement of electronic structure of (TaSe$_4$)$_2$I at 285~K.
The single crystal is with in-situ reproducibly cleaved to expose (110) surface due to its quasi-1D nature~\cite{SMtiii}.
The stripe-like constant energy contour near Fermi level in Fig.~\ref{fig3}(a) indicates a strong nesting, which contributes to the CDW transition below 263~K~\cite{tournier2013PRL,wang1983zz,maki1983charge}.
The ARPES spectra along cuts (1, 2, 3), marked as black dashed lines in Fig.~\ref{fig3}(a), are presented in Figs.~\ref{fig3}(c)-\ref{fig3}(e), respectively.
A V-shaped dispersion in cut1 in Fig.~\ref{fig3}(c) indicates a linearly dispersing band structure near $W_1$.
In addition, we find the dispersions along cut2 and cut3 exhibit a quadratic characteristic in the vicinity of $W_2$ and $W_1$, respectively.
The dashed red lines in Figs.~\ref{fig3}(c)-\ref{fig3}(e) are the calculated band structures along certain direction, and the measured ARPES spectra show good agreement with our calculations, supporting that $\mathrm{(\mathrm{Ta}Se_{4})_{2}I}$ is a type-III Weyl semimetal at room temperature.
In addition, we calculate the energy bands of three types of Weyl semimetals in the presence of homogeneous magnetic fields along the $z$ direction and find that the type-III Weyl semimetals exhibit a unique Landau level's turn-up due to the dominant quadratic tilting term. This peculiar Landau level spectra can be directly probed by magneto-optic experiments and act as a smoking gun signature of type-III Weyl fermions~\cite{SMtiii}.
%
\begin{figure}[t]
\includegraphics[width=8.2cm]{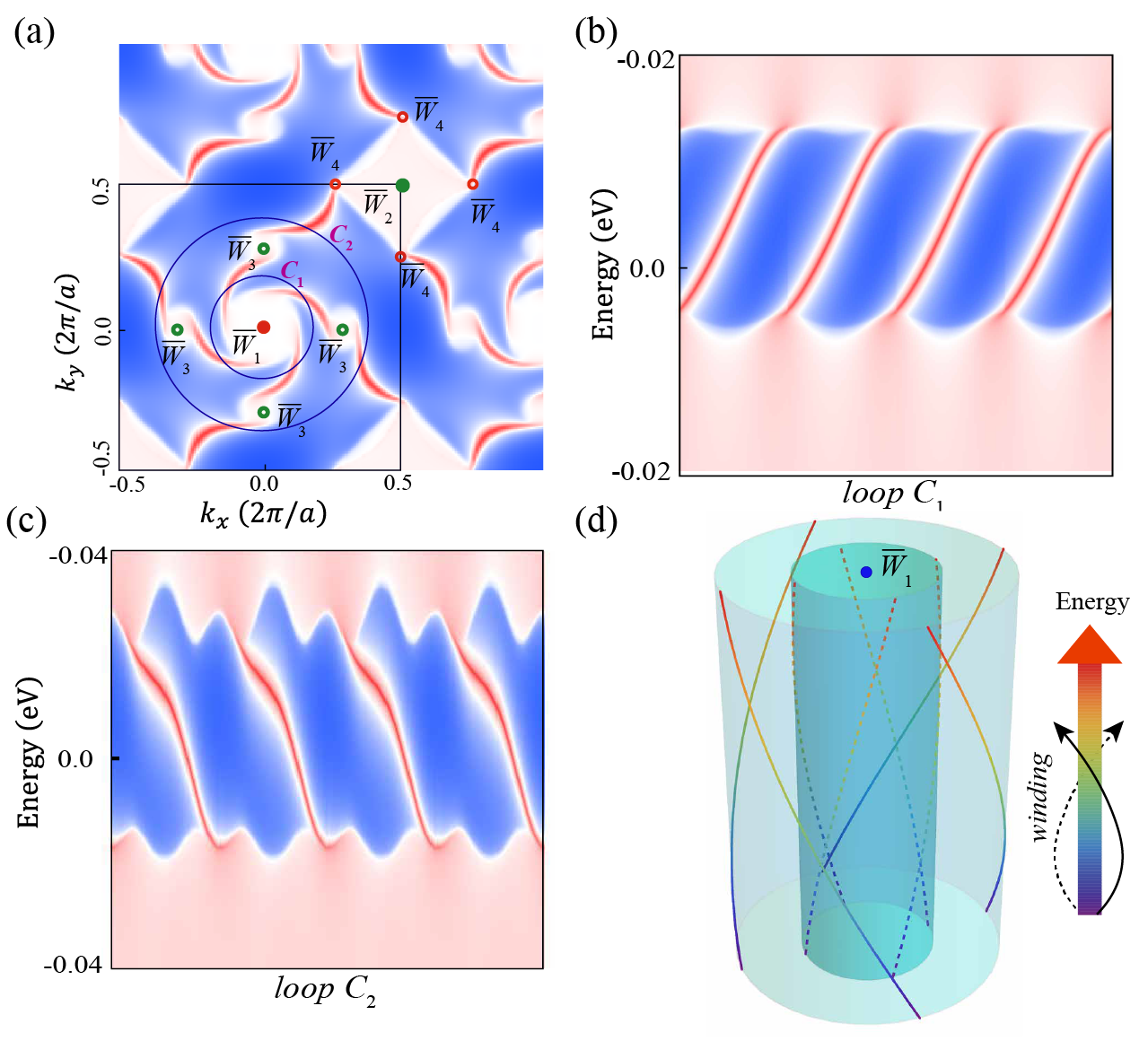}
\caption{(Color online) Calculated electronic structures on (001) surface. (a) Four-fold helicoidal surface arcs at Fermi level.
(b) Perspective plot of the dispersion of chiral edge modes from (c) and (d), respectively.
The winding of the chiral modes around $W_1$ as a function of energy suggests that the Fermi arcs have a four-fold helicoid structure.
(c)-(d) Surface LDOS along the loop $C_{1}$ and $C_{2}$ in (a), respectively.
\label{fig4}}
\end{figure}

Interestingly, our calculations reveal more exotic features of the Fermi arc surface states on (001) plane.
As shown in Fig.~\ref{fig4}(a), two type-III double WPs ($W_{2}$) above the Fermi energy are projected into the corner ($\overline{W}_{2}$) of the surface BZ, forming a hole pocket with $\chi=4$, while the other type-III double WPs ($W_{1}$) below the Fermi energy are projected into the center ($\overline{W}_{1}$) of surface BZ, forming an electron pocket with $\chi=-4$.
Meanwhile, four pairs of type-I single WPs ($W_{3}$) related by $C_4$ rotation are projected into four $\overline{W}_{3}$ ($\chi=2$) surrounding the $\overline{W}_{1}$ point.
Similarly, four pairs of type-II single WPs ($W_{4}$) are projected into four $\overline{W}_{4}$ surrounding $\overline{W}_{2}$ point, forming four hole pockets with $\chi=-2$.
So the four hole pockets of $\overline{W}_{4}$ merge together with the hole pocket of $\overline{W}_{2}$, forming a larger square-like hole pocket with $\chi=-4$.
One can see that four branches of Fermi arcs stem from $\overline{W}_{1}$, pass through $\overline{W}_{3}$, connect to $\overline{W}_{4}$ and finally merge into $\overline{W}_{2}$.
Those long Fermi arcs that connect the $\overline{W}_{1}$ and $\overline{W}_{2}$ demonstrate a four-fold helicoidal nature~\cite{fang2016np,sanchez2019topological,rao2019new}.
To visualize the helicoidal surface states, we calculate the surface local density of states (LDOS) along two clockwise loops ($C_{1,2}$) centered at $\overline{W}_{1}$, as shown in Figs.~\ref{fig4}(b) and \ref{fig4}(c).
For the loop $C_1$, the four right-moving chiral edge modes appear insides the band gap in line with the chirality of $\overline{W}_{1}$.
Since the loop $C_2$ encompass both $\overline{W}_{3}$ and $\overline{W}_{1}$, the four edge modes instead have opposite chirality.
The four-fold right-handed and left-handed spiral surface state along the loop $C_1$ and $C_2$ are schematically demonstrated in Fig.~\ref{fig4}(d).
Note that since the naturally cleaved surface of $\mathrm{(\mathrm{Ta}Se_{4})_{2}I}$ is the (110) surface, the such novel topological surface states that exist on the (001) surface are therefore absent from our ARPES measurements. Instead, the signature of topological surface state could be accessible to transport measurement of micro-fabricated nano-scaled samples~\cite{wang2017quantum}.

%

\textit{Strain induced transitions.--}We stress that a pair of double WPs with $\chi=2$ locate at the corner of BZ $P$ and $-P$, whereas the other pair with $\chi=-2$ locates on the $C_4$ invariant axis thus can move along the ${\Gamma}$-Z path.
Thus, there is no loop available for annihilation of the two WPs with opposite chirality as long as the crystal symmetry is preserved.
It implies that these type-III double WPs in $\mathrm{(\mathrm{Ta}Se_{4})_{2}I}$ exhibit remarkable robustness against external perturbations that respect crystal symmetries.
\begin{figure}[t]
\includegraphics[width=8.0cm]{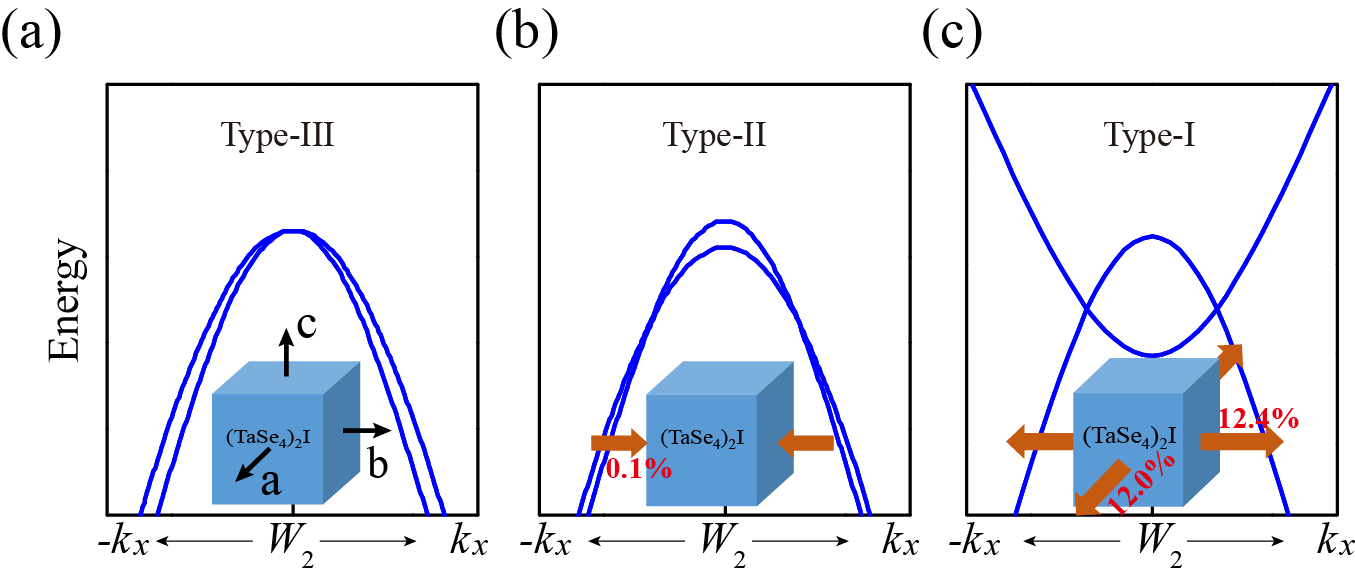}
\caption{(Color online) The strain tuned  transitions in $\mathrm{(\mathrm{Ta}Se_{4})_{2}I}$ among type-III, type-II and type-I Weyl semimetals.
(a) The energy dispersion along the $k_{x}$ direction near $W_{2}$ (red cut of the 3D energy dispersion in Fig.~\ref{fig2}(f)).
(b) The energy dispersion around type-II WP under 0.1$\%$ uniaxial strain (0.02~GPa) along the (100) direction.
(c) The energy dispersion around type-I WP in the presence of 12.0$\%$(1.20~GPa) and 12.4$\%$(1.22~GPa) biaxial tensile strain along the a and c axes.
\label{fig5}}
\end{figure}

Strains have proven to be a clear and effective way to tune the electronic structure and the relevant physical properties.
We now investigate the impacts of strains that break crystal symmetries on the electronic properties in $\mathrm{(\mathrm{Ta}Se_{4})_{2}I}$.
Since the type-III double WPs are protected by $C_4$ or $C_4T$, applying a symmetry-breaking uniaxial strain along (100)/(010) direction could split a type-III double WP with $\chi=2$ into two type-II WPs with unit chiral charge, accompanied with Lifshitz transition of the Fermi surface as shown in Figs.~\ref{fig5}(a)-\ref{fig5}(b).
We also estimate that about 0.1$\%$ (0.02~GPa) compressive strain along $a$ axis could induce such topological transitions.
Furthermore, imposing an extra biaxial in-plane strain, the type-II single WP will further transform into type-I single WP along with a second Lifshitz transition of the Fermi surface, as shown in Fig.~\ref{fig5}(c).
Therefore, these three types of Weyl fermions can be mutually transformed through strain engineering in $\mathrm{(\mathrm{Ta}Se_{4})_{2}I}$, which provides a tunable platform for studying exotic properties of various Weyl semimetals.
In experiments, these transitions among different Weyl fermions can be induced via proper external pressure and the relevant Lifshitz transitions of Fermi surface as well as the corresponding surface states offer a measurable signal that can be probed by magnetic oscillation experiments~\cite{shoenberg1984}, ARPES~\cite{LvBQ2015PRX,Xu2015Science} and magneto-transport~\cite{Kim2013PRL,Huang2015PRX,Xiong2015science,li2015NC,LiH2016NC,LiQ2016NP}.
It is worth noting that $\mathrm{(\mathrm{Ta}Se_{4})_{2}I}$ could support a new solid state phase combining long-range ordered but short-range disordered structure under pressure (about 20 GPa) as well as the anomalous two-step superconducting transition~\cite{AnC2020SC}.

In summary, we propose the new type-III Weyl semimetal whose Fermi surface that consists of two touched electron or hole pockets and first provide experimental evidence of such novel fermion in $\mathrm{(\mathrm{Ta}Se_{4})_{2}I}$ by means of first principle calculations and ARPES experiments.
Two pairs of type-III Weyl points emerge due to the over-tilted quadratic terms and are protected by $C_4$ and $C_4T$ symmetries, respectively.
We find that the strains that break $C_4$ symmetry trigger the  transitions from type-III to type-II, and to type-I WPs along with Lifshitz transitions of the Fermi surface.
In addition, a four-fold helicoidal surface state is predicted on (001) surface with long Fermi arcs.
Therefore, our work reports the first material realization of multi-Weyl semimetal and also provides a promising platform to further study the novel physics of type-III Weyl semimetals.

\begin{acknowledgments}
We are grateful to Zhi-Ming Yu for fruitful discussions. The work at BIT is supported by the National Key R\&D Program of China (Grant No. 2016YFA0300600), the National Natural Science Foundation of China (Grants Nos. 11734003), the Strategic Priority Research Program of Chinese Academy of Sciences (Grant No. XDB30000000).
K.D. and S.Z. are supported by the National Natural Science Foundation of China (Grants No. 11725418).
J.Z. is supported by the High Magnetic Field Laboratory of Anhui Province.
\end{acknowledgments}%

%

\end{document}